\documentclass[twocolumn, 5p]{elsarticle}

\makeatletter
\def\ps@pprintTitle{%
  \let\@oddhead\@empty
  \let\@evenhead\@empty
  \def\@oddfoot{\reset@font\hfil\thepage\hfil}
  \let\@evenfoot\@oddfoot
}
\makeatother

\usepackage{amsmath, amssymb, graphicx, lmodern, physics, xcolor}
\usepackage[colorlinks=true, urlcolor=blue, anchorcolor=blue, citecolor=blue, filecolor=blue, linkcolor=blue, menucolor=blue, linktocpage=true, pdfproducer=medialab, pdfa=true]{hyperref}
\usepackage{varioref, cleveref}

\usepackage{etoolbox}
\apptocmd{\sloppy}{\hbadness 10000\relax}{}{}

\renewcommand{\d}{\partial}

\newcommand\mperiod[1][\rlap]{#1{\ .}}	
\newcommand\mcomma[1][\rlap]{#1{\ ,}}

\crefname{table}{table}{tables}
\Crefname{table}{Table}{Tables}
\crefname{figure}{figure}{figures}
\Crefname{figure}{Figure}{Figures}

\newenvironment{eqaed}
    {\begin{equation}
    \begin{aligned}
    }
    { 
    \end{aligned}
    \end{equation}
    \ignorespacesafterend
    }

\begin{document}

\title{\textbf{On New Vacua of non-Supersymmetric Strings}}
\author[1]{Salvatore Raucci}
\ead{salvatore.rauci@sns.it}
\affiliation[1]{organization={Scuola Normale Superiore and INFN},
addressline={Piazza dei Cavalieri 7},
postcode={56126},
city={Pisa},
country={Italy}}

\begin{abstract}
We describe two new types of vacua for ten-dimensional string models without supersymmetry, in which dilaton potentials are compensated by constant electric or magnetic fields.
These arise from $\alpha '$ corrections to the equations of motion, and we comment on the reliability of this expansion.
We identify explicitly unstable singlet scalar perturbations in the orientifold vacua, and we argue that in both cases additional instabilities are induced by non-abelian couplings.
\end{abstract}

\maketitle

\section{Introduction and summary}

The search for string vacua without supersymmetry remains a topic of utmost importance. In this work, we shall start directly from the ten-dimensional tachyon-free models without supersymmetry and look for classical vacua. 
The models of interest are the heterotic $\text{SO}(16)\times\text{SO}(16)$ string~\cite{AlvarezGaume:1986jb, Dixon:1986iz} and the U(32) orientifold~\cite{Sagnotti:1987tw, Pradisi:1988xd, Horava:1989vt, Horava:1989ga, Bianchi:1990yu, Bianchi:1990tb, Bianchi:1991eu, Sagnotti:1992qw} model~\cite{Sagnotti:1995ga, Sagnotti:1996qj}, in which supersymmetry is absent, and Sugimoto's USp(32) orientifold~\cite{Sugimoto:1999tx}, the simplest setting for ``brane supersymmetry breaking''~\cite{Antoniadis:1999xk, Angelantonj:1999jh, Aldazabal:1999jr, Angelantonj:1999ms} with non-linearly realized supersymmetry. 
Our understanding of all these models is challenged by a non-trivial spacetime back-reaction, which results in the presence of a runaway ``tadpole potential'' for the dilaton. 
The leading effect of supersymmetry breaking on the low-energy action is captured by a scalar potential
\begin{eqaed}\label{eq:tadpole_contribution}
    V(\phi) = T e^{\gamma\phi}\mcomma 
\end{eqaed}
with model-dependent $T$ and $\gamma$. In particular, in the Einstein frame $\gamma=\frac{3}{2}$ for the two orientifold models, while $\gamma=\frac{5}{2}$ for the heterotic model. We refer the reader to the excellent reviews of~\cite{Angelantonj:2002ct, Mourad:2017rrl, Basile:2020xwi, Basile:2021mkd, Basile:2021vxh} for introductions from a string theory perspective.

The tadpole potential forbids a ten-dimensional Minkowski vacuum, leading, in the simplest setting, to the Dudas-Mourad compactification~\cite{Dudas:2000ff}, with curvature singularities and strong coupling regions. However, it can balance the effects of fluxes and result in a stabilized dilaton. 
In fact, in~\cite{Gubser:2001zr, Mourad:2016xbk} the authors used R-R fluxes to yield constant dilaton profiles. In a similar fashion, in this paper we shall exploit the presence of gauge fields in the three models of interest, turning on non-zero vacuum values for U(1) abelian gauge fields\footnote{The gauge algebra in the U(32) model is actually that of SU(32), because the remaining U(1) is anomalous~\cite{Sagnotti:1996qj} and cannot be used as the background U(1).} in their Cartan subalgebras.
This choice adapts to all models, regardless of their gauge symmetries, and it is compatible with the simple choices of manifolds under scrutiny.

We can already anticipate an issue of the construction, the need to mix different $\alpha '$ corrections. In \cref{eq:tadpole_contribution} one can argue, for purely dimensional reasons, that there is a factor $1/\alpha '$ in the definition of $T$. Consequently, any solution will necessarily scale with $\alpha '$ in a non-trivial fashion. In the vacua of~\cite{Dudas:2000ff} with reduced isometry, a factor $\sqrt{\alpha '}$ accompanies a spacelike (or timelike for the cosmological solutions) coordinate.
In~\cite{Gubser:2001zr, Mourad:2016xbk} and here, R-R or gauge fields must scale with a negative power of $\alpha '$ to yield constant dilaton profiles. 

In \cref{sec:orientifold_models} we discuss these settings in the two orientifold models. After presenting the solution, we investigate its perturbative stability and find explicitly unstable singlet scalar modes. In \cref{sec:heterotic_models} we repeat the analysis in the heterotic $\text{SO}(16)\times\text{SO}(16)$ model, and find a ``dual'' solution with no explicit unstable scalar modes. However, we argue that both classes of vacua suffer from another source of instability. In \cref{sec:heterotic_higher_der} we reconsider the general strategy, highlighting some problems emerging from the $\alpha'$-mixing. In fact, gauge kinetic terms arise from a double expansion in both $\alpha '$ and spacetime derivatives, and we explain when this expansion can be reliable.

\section{Electrovac solutions for the USp(32) and U(32) orientifold models}\label{sec:orientifold_models}

Let us begin by recalling the relevant portions of the low-energy effective actions that are common to the USp(32) and U(32) orientifolds, which will play a role in our analysis. The contributions of interest arise from gravity, dilaton and the gauge kinetic terms, which are accompanied by the dilaton dependent factor $e^{-\phi}$ in the string frame. The dilaton tadpole potential scales proportionally to $e^{-\phi}$ in both models, so that the relevant action is
\begin{eqaed}\label{eq:tadpole_orientifold_action_string_frame}
    \int \! \sqrt{-g} \, \left[e^{-2\phi}\left(R+4(\d\phi)^2\right)-\frac{\alpha'}{4} e^{-\phi}\tr F^2 - T  \,  e^{-\phi}\right]\mcomma
\end{eqaed}
where $T$ is positive but different in the two cases.

The metric ansatz that we shall consider involves a product of two maximally symmetric spaces
\begin{eqaed}
    ds^2 =\lambda_{\mu\nu} \, dx^\mu dx^\nu + \gamma_{ij} \, dy^i dy^j\mperiod
\end{eqaed}
This assumption, in the presence of constant dilaton profiles, leads to
\begin{eqaed}\label{eq:maxwell_propto_metric}
    F_{MA}{F_N}^A \propto g_{MN} \mcomma
\end{eqaed}
where the proportionality constant can also vanish.
The simplest available choice to satisfy this condition is to employ a U(1) field strength in the Cartan subalgebra of USp(32) or SU(32).
Since $F_{MN}$ is a two-tensor, these types of vacua must involve a two-dimensional space, and the equations of motion
\begin{eqaed}\label{eq:orientifold_eom_u(1)}
    & \frac{\alpha'}{8}F^2+3T = 0 \mcomma \qquad \nabla_M {F}^{MN} =0 \mcomma  \\
    & R_{MN}-\frac{1}{2}T  \, e^{\phi_0}g_{MN}-\frac{\alpha'}{4} \,  e^{\phi_0}F_{MA}{F_N}^A = 0\mcomma
\end{eqaed}
call for a two-dimensional spacetime with an electric background, so that $F_{\mu\nu} = \epsilon_{\mu\nu} f$.
In fact, writing the field strength in terms of the vielbein as
\begin{eqaed}\label{eq:constant_electric_and_magnetic_field}
    F_{MN}=E_{i}(e_M^0 e_N^i-e_N^0 e_M^i)+H_{ij}(e_M^i e_N^j-e_N^i e_M^j)\mcomma 
\end{eqaed}
one needs a vanishing $H_{ij}$ and a constant $E_i$ along the spatial direction of the two-dimensional spacetime in order to solve \cref{eq:orientifold_eom_u(1)}. For this reason, we call this an ``electrovac'', a terminology that was already used in~\cite{Freedman:1983xa, Antoniadis:1989mn} for the supersymmetric case. 

Combining the preceding considerations leads to
\begin{eqaed}\label{eq:electric_vacuum_string_frame}
    & R_{\mu\nu}  = - \frac{5}{2} T  \,  e^{\phi_0}\lambda_{\mu\nu} \mcomma \qquad R_{mn} = \frac{1}{2} T \,  e^{\phi_0} \gamma_{mn}\mcomma \\ & \alpha' f^2 =12T  \mperiod
\end{eqaed}
The two-dimensional spacetime is thus an $AdS_2$, while the internal manifold is an Einstein manifold with positive curvature. In what follows, we shall assume it to be an $S^8$ with the round metric.

As anticipated, in order to compensate the dilaton potential, the electric field must scale as $f\sim 1/\alpha'$.

\subsection{The issue of perturbative stability}\label{sec:electr_instabilities}

We now investigate the perturbative stability of this new solution. We turn to the Einstein frame, where the linearized equations afford a clearer interpretation.

At the linearized level, there are no mixings between non-abelian perturbations, which do not contribute to the background profile, and the others. 
For the time being, let us therefore confine our attention to perturbations of the dilaton, the metric, and the U(1) gauge field.
We denote these as $\varphi, h_{MN}$ and $a_M/f$, where now $f$ is the parameter that enters the background Einstein-frame electric field, $\alpha ' f^2=12 \, e^{\phi_0} \, T$. These will depend on the $AdS_2$ coordinates $x^\mu$ and the $S^8$ coordinates $y^i$, and will be expanded in spherical harmonics.
In what follows, covariant derivatives and curvature tensors always refer to the background.

The metric equations of motion become
\begin{eqaed}\label{eq:linearization_electric_metric}
    \Box h_{MN} &=-\nabla_M \nabla_N h +2\nabla_{(M} \nabla^R h_{N)R}+\\
        & +2{R^B}_{(N} h_{M)B}-2{R^B}_{MAN}{h_B}^A+ \\
            & -R_{MN}\varphi+\frac{1}{2}T \,  e^{\frac{3}{2}\phi_0}{g}_{MN}\varphi-T \,  e^{\frac{3}{2}\phi_0}h_{MN}+ \\
                &-\frac{\alpha'}{f}  \,  e^{\frac{1}{2}\phi_0}\bigg[{F_M}^A \, \nabla_{[N}a_{A]}+ {{F}_N}^A \, \nabla_{[M}a_{A]}+\\
                    & -\frac{f}{2}{F}_{MA}{F}_{NP}h^{PA}\bigg]+\\
                        &+{g}_{MN}\bigg[\frac{\alpha'}{8f} \,  e^{\frac{1}{2}\phi_0}{F}_{AB} \, \nabla^{[A} a^{B]}+\frac{1}{8}T \,  e^{\frac{3}{2}\phi_0}h+ \\
                            & -\frac{3}{4}T \, e^{\frac{3}{2}\phi_0}\varphi-\frac{1}{4}R_{AB}h^{AB}\bigg]\mcomma
\end{eqaed}
and $h_{MN}$ yields, in principle, tensor, vector and scalar modes in $AdS_2$.

The dilaton equation becomes
\begin{eqaed}\label{eq:linearization_electric_dilaton}
    \Box\varphi &=  \frac{3}{2}T  \,  e^{\frac{3}{2} \phi_0}\varphi + \frac{\alpha'}{4f} \, e^{\frac{1}{2}\phi_0} F_{AB} \nabla^Aa^B+ \\ & -\frac{1}{2}R_{MN}h^{MN} + \frac{1}{4} T  \,  e^{\frac{3}{2}\phi_0} h  \mcomma
\end{eqaed}
and $\varphi$ only contributes to scalar perturbations.

The remaining linearized equations for the U(1) gauge field read
\begin{eqaed}\label{eq:linearization_electric_gauge}
    \Box\frac{a^N}{f} &=\nabla_M \nabla^N \frac{a^M}{f}+{{F}_P}^N \nabla_M h^{MP}+\\
        & + {{F}^M}_P \nabla_M h^{NP}-\frac{1}{2}{F}^{MN}\nabla_M (h+\varphi)\mcomma
\end{eqaed}
and describe, in principle, both vector and scalar modes.

We can now study the squared masses for the allowed modes arising from the above linearized equations, looking for violations of the Breitenlohner–Freedman (B-F) bound~\cite{Breitenlohner:1982jf}. Before presenting the results, it is convenient to simplify our notation, letting
\begin{eqaed}\label{eq:units_orientifold}
    \tau\equiv T  \, e^{\frac{3}{2}\phi_0}\mcomma \qquad L\equiv \frac{l(l+7)}{14}\mperiod
\end{eqaed}
$L$ enters the equations through the eigenvalues of the Laplacian on the internal sphere, while $\tau$ sets our units.
We shall also denote the internal Laplacian, in a way that should not create any confusion, as $\Box_8$.

\subsection{Tensor and vector perturbations}\label{sec:eletro_instabilities_tensor_vector}

Tensor modes in $AdS$ would arise from \cref{eq:linearization_electric_metric}, describing a massless graviton and a tower of KK excitations, with
\begin{eqaed}\label{eq:electro_tensor_modes}
    \Box h_{\mu\nu}=-\frac{2}{l_{AdS}^2} h_{\mu\nu}\mperiod
\end{eqaed}
However, in two dimensions, these modes are pure gauge. 
The same happens for the vector modes, for which \cref{eq:linearization_electric_gauge} with $N=\mu$ gives Maxwell's equations in $AdS$
\begin{eqaed}\label{eq:maxwell_AdS}
    \Box a_{\mu}=\nabla^\nu\nabla_\mu a_\nu \mcomma
\end{eqaed}
and for the graviphoton $h_{\mu i}$ from \cref{eq:linearization_electric_metric}, taking into account that it only generates $l\geq1$ modes.

\subsection{Scalar perturbations}\label{sec:eletro_instabilities_scalar}

Let us begin with the transverse traceless modes $h_{ij}$, which are tensors with respect to the internal rotation group. These are stable, since from \cref{eq:linearization_electric_metric}
\begin{eqaed}\label{eq:orientifold_electric_tensor_pert}
   \Box_2 h_{ij}  =L\tau  h_{ij} \mcomma
\end{eqaed}
where $\Box_2$ denotes the $AdS_2$ d'Alembertian.

In addition, there are internal vectors from $h_{\mu i}$ and the internal components of the gauge field. Letting
\begin{eqaed}\label{eq:electro_vector_pert_gauge}
    a_i\equiv \frac{1}{\alpha'}\Omega_i e^{-\frac{1}{2}\phi_0}\mcomma \qquad  h_{\mu i}=\tau^{-1}  \epsilon_{\mu\nu}\nabla^\nu V_i\mcomma 
\end{eqaed}
these mix according to
\begin{eqaed}\label{eq:electro_vector_mass_matrix}
    \Box_2 \begin{pmatrix}V_i \\ \Omega_i \end{pmatrix} = 
    \begin{pmatrix}L-\frac{4}{7} & \frac{1}{2} \\ 12L-\frac{48}{7} & L+\frac{45}{7} \end{pmatrix}\tau \begin{pmatrix}V_i \\ \Omega_i \end{pmatrix}\mperiod
\end{eqaed}
The mass matrix in \cref{eq:electro_vector_mass_matrix} has a vanishing eigenvalue for $l=1$, while the others are positive. Therefore, this sector contains no unstable modes.

Singlet scalar perturbations take the form
\begin{eqaed}\label{eq:orientifold_electric_scalar_pert_definitions}
    h_{\mu\nu} & = A \lambda_{\mu\nu} \mcomma \qquad h_{ij} = C \gamma_{ij} \mcomma \qquad h_{\mu i}  = \nabla_\mu \nabla_i D\mcomma \\  a_\mu & = \frac{1}{\alpha'}e^{-\frac{1}{2}\phi_0}  \epsilon_{\mu\nu}\nabla^\nu \Omega \mcomma
\end{eqaed}
where we excluded some contributions that can be gauged away. 
The $\mu\nu$ and $ij$ components of \cref{eq:linearization_electric_metric} involve two different tensor structures, $\sim \lambda_{\mu\nu}$ and $\sim \nabla_\mu \nabla_\nu$, which must be dealt with separately, as in~\cite{Basile:2018irz}.
Then, the linearized equations lead to
\begin{eqaed}\label{eq:orientifold_electric_scalar_pert}
    &\Box \Omega  = -12\tau \left[-\frac{1}{2}\varphi+\Box_8 D + A - 4 C\right]\mcomma \\
    &\Box\varphi  = \frac{3}{2} \tau \varphi+3\tau A +\frac{1}{4}\Box_2 \Omega \mcomma \\
    &\Box A  = \frac{9}{4} \tau \varphi-\frac{11}{2} \tau  A -\frac{7}{8}\Box_2 \Omega \mcomma \\
    &\Box_8 D  = 4C\mcomma \\
    &\Box C  = - \tau C+\frac{1}{8}\Box_2 \Omega +\frac{3}{2} \tau A-\frac{3}{4} \tau \varphi\mcomma\\
    &\Box_2 D  = A+3C \mcomma \\
    &0  =  A+7C+ \tau D+\frac{1}{2}\Omega \mperiod
\end{eqaed}

If $l=0$, $D$ is absent, $C=0$ and $\Omega$ decouples, corresponding to pure gauge perturbations. One is then left with a mass matrix for the fields $A$ and $\varphi$, whose eigenvalues are both positive.

The case $l=1$ implies $L=\frac{4}{7}$ and $\Omega=-2A$, and leads to a mass matrix for the modes with two positive and one negative eigenvalue, equal to  $-\frac{3}{7}\tau$. This is still compatible with the B-F bound $-\frac{5}{8}\tau$, and these perturbations are stable.

In the general case $l>1$, $C$ and $D$ can be expressed in terms of the other fields and the remaining modes are
\begin{eqaed}\label{eq:electro_scalar_instab}
    \Box_2 \begin{pmatrix}A \\ \Omega \\ \varphi \end{pmatrix} = 
    \begin{pmatrix}
    L+5 & -\frac{7}{8}L & -3 \\ 
    -12 & L & 6 \\
    0 & \frac{1}{4}L & L+3 \end{pmatrix}\tau \begin{pmatrix}A \\ \Omega \\ \varphi \end{pmatrix}\mperiod
\end{eqaed}
Only one of the eigenvalues can be negative.
Taking into account the B-F bound for scalar modes, one can see that the solution has three unstable sets of scalar modes, with $l=2,3,4$.

In the above analysis, the internal manifold is a sphere with the round metric, but any Einstein manifold with appropriate curvature would yield a vacuum. In our notation, $L$ was chosen so that it denotes sphere eigenvalues in units of $\tau$, and a different Einstein manifold would correspond to different values of $L$. Stability can thus be rephrased as a property of the new Laplacian eigenvalues. In particular, a large enough Laplacian gap, corresponding to $L\gtrapprox4$, would remove the scalar instability from \cref{eq:electro_scalar_instab}. We have no argument at present to exclude this option, but we have no viable example.

\section[Magnetovac solutions for the heterotic SO(16) x SO(16) model]{Magnetovac solutions for the heterotic $\text{SO}(16)\times\text{SO}(16)$ model}\label{sec:heterotic_models}

In this section, we address a similar construction for the heterotic non-supersymmetric $\text{SO}(16)\times\text{SO}(16)$ model. The effective action in the string frame has in this case a cosmological constant term with $\Lambda>0$, and the absence of a dilaton coupling reflects the one-loop origin of this contribution. On the other hand, gauge fields are accompanied by a dilaton factor $e^{-2\phi}$, which signals their closed string origin, so that the two-derivative action, up to terms of order $\alpha'$, is
\begin{eqaed}\label{eq:heterotic_action_string_frame}
    \int \! \sqrt{g} \, \left[e^{-2\phi}\left(R+4(\d\phi)^2-\frac{\alpha'}{4}\tr F^2\right)-\Lambda \right]\mperiod
\end{eqaed}

One can retrace the steps of \cref{sec:orientifold_models}, searching for solutions with constant dilaton profiles and vacuum values for U(1) gauge fields.
Now the equations of motion demand $F^2>0$, and thus lead to a magnetic background confined to a two-dimensional internal manifold. This corresponds to a non-vanishing $H_{89}$ in \cref{eq:constant_electric_and_magnetic_field}.
One thus obtains an $AdS_8\times S^2$ solution, with $F_{mn}=\epsilon_{mn} f$, which is characterized by
\begin{eqaed}\label{eq:magnetic_vacuum_string_frame}
   &  R_{\mu\nu} = -\frac{1}{2}\Lambda \,  e^{2\phi_0} \, \lambda_{\mu\nu}\mcomma \qquad R_{mn} = \frac{9}{2}\Lambda  \, e^{2\phi_0} \,  \gamma_{mn}\mcomma\\
   & \alpha' f^2 = 20 \Lambda \, e^{2\phi_0} 
\end{eqaed}
in the string frame.

As in \cref{eq:electric_vacuum_string_frame}, the gauge field strength must be proportional to $1/\alpha'$, and the tadpole contribution is compensated by an $\alpha'$ correction. Using flux quantization on the compact sphere, one could trade $\phi_0$ for the U(1) magnetic flux $N_m\propto e^{-\phi_0}$, but in what follows we shall keep $\phi_0$.

\subsection{The issue of perturbative stability}\label{sec:magnet_instabilities}

We now investigate the perturbative stability, again in the Einstein frame. 
One can linearize the equations of motion using the same notation as in the orientifold case. Again, we turn on gauge field deformations along the same U(1) that hosts the vacuum profile.

The gravitational equations of motion become
\begin{eqaed}\label{eq:linearization_magnetic_metric}
    \Box h_{MN} &=-\nabla_M \nabla_N h +2\nabla_{(M} \nabla^R h_{N)R}+\\
        & +2{R^B}_{(N} h_{M)B}-2{R^B}_{MAN}{h_B}^A+\\
            &+R_{MN}\varphi+\frac{1}{2}\Lambda \,  e^{\frac{5}{2} \phi_0}{g}_{MN}\varphi+\Lambda  \, e^{\frac{5}{2} \phi_0}h_{MN}+ \\
                &-\frac{\alpha'}{f} e^{-\frac{1}{2}\phi_0}\bigg[{{F}_M}^A \nabla_{[N} a_{A]} + {{F}_{N}}^A \nabla_{[M} a_{A]}+ \\
                    &  - \frac{f}{2}{F}_{MA}{F}_{NP}h^{PA}\bigg]+ \\
                        & +{g}_{MN}\bigg[\frac{\alpha'}{8f} \, e^{-\frac{1}{2}\phi_0}{F}_{AB} \, \nabla^{[A} a^{B]}+ \\
                            & -\frac{1}{8}\Lambda \,  e^{\frac{5}{2} \phi_0}h-\frac{5}{4}\Lambda  \, e^{\frac{5}{2} \phi_0}\varphi-\frac{1}{4}R_{AP}h^{AP}\bigg]\mcomma
\end{eqaed}
where $h_{MN}$ generates all types of perturbations.

The linearized dilaton equation is
\begin{eqaed}\label{eq:linearization_magnetic_dilaton}
    \Box\varphi &= \frac{15}{2}\Lambda  \, e^{\frac{5}{2} \phi_0}\varphi - \frac{\alpha'}{4f} \, e^{-\frac{1}{2}\phi_0} F_{AB}\nabla^A a^B+ \\
    & + \frac{1}{2}R_{MN}h^{MN}+\frac{1}{4}\Lambda \,  e^{\frac{5}{2} \phi_0} h  \mcomma
\end{eqaed}
and $\varphi$ clearly contributes to scalar perturbations.

The gauge field, whose linearized equations become
\begin{eqaed}\label{eq:linearization_magnetic_vacua}
    \Box\frac{a^N}{f} &=\nabla_M \nabla^N \frac{a^M}{f}+{{F}_P}^N \nabla_M h^{MP}+\\
    & + {{F}^M}_P \nabla_M h^{NP}-\frac{1}{2}{F}^{MN}\nabla_M (h-\varphi)\mcomma
\end{eqaed}
describes both vector and scalar modes.

As in \cref{sec:electr_instabilities}, one must study the squared masses of the modes arising from these equations, and compare them with the B-F bounds. Let us again simplify our notation, letting
\begin{eqaed}\label{eq:units_heterotic}
    \tau\equiv \Lambda  \,  e^{\frac{5}{2}\phi_0}\mcomma \qquad L\equiv \frac{9l(l+1)}{2}\mperiod
\end{eqaed}
This is the counterpart of \cref{eq:units_orientifold}, and $L$ describes the eigenvalues of the Laplacian on the internal sphere in units of $\tau$.

\subsection{Tensor, vector and non-singlet scalar perturbations}\label{sec:magneto_instabilities_tensor_vector}

Tensor modes in $AdS_8$ arise from \cref{eq:linearization_magnetic_metric}. They behave as in \cref{eq:electro_tensor_modes} and describe a massless graviton for $l=0$ and a tower of KK excitations for $l\neq 0$.

Vector modes arise from divergence-free metric perturbations $h_{\mu i}$ and gauge field perturbations $a_\mu$. Letting
\begin{eqaed}
    a_{\mu}\equiv \frac{1}{\alpha'}\Omega_\mu e^{\frac{1}{2}\phi_0}  \mcomma \qquad h_{\mu i}=\tau^{-1} \epsilon_{ji}\nabla^j V_\mu\mcomma
\end{eqaed}
one is left with
\begin{eqaed}\label{eq:heterotic_vector_modes}
    \Box_8 \begin{pmatrix}V_\mu \\ \Omega_\mu \end{pmatrix} =
    \begin{pmatrix}L+\frac{1}{2} & -\frac{1}{2}  \\ -20 {L} & L-\frac{1}{2} \end{pmatrix}\tau \begin{pmatrix}V_\mu \\ \Omega_\mu \end{pmatrix}\mperiod
\end{eqaed}
When $l=0$, only the $a_\mu$ modes are present, subject to the massless $AdS$ Maxwell equations, as in \cref{eq:maxwell_AdS}. When $l\neq0$, the mass matrix in \cref{eq:heterotic_vector_modes} describes a triplet of massless vectors, corresponding to one of the $l=1$ eigenvalues, together with infinitely many massive modes.

\subsection{Scalar perturbations}\label{sec:magneto_instabilities_scalar}

Tensor modes with respect to the internal rotation group arise from $h_{ij}$. From \cref{eq:linearization_magnetic_metric} one obtains the counterpart of \cref{eq:orientifold_electric_tensor_pert}, now with $\Box_2\to \Box_8$. Both these and the internal vector modes $a_i$ from \cref{eq:linearization_magnetic_vacua} can be shown to be stable.  

Let us complete the analysis by considering singlet scalar perturbations. Echoing \cref{eq:orientifold_electric_scalar_pert_definitions}, these take the form
\begin{eqaed}\label{eq:heterotic_magnetic_scalar_pert_definitions}
    h_{\mu\nu} &= A \lambda_{\mu\nu} \mcomma \qquad h_{ij}  = C \gamma_{ij} \mcomma \qquad h_{\mu i}  = \nabla_\mu \nabla_i D\mcomma \\    
    a_i & =\frac{1}{\alpha'} e^{\frac{1}{2}\phi_0} \epsilon_{ij}\nabla^j \Omega \mcomma
\end{eqaed}
where we have neglected contributions that are pure gauge. Again, there are two different tensor structures from \cref{eq:linearization_magnetic_metric}, which must be dealt with separately. One obtains
\begin{eqaed}\label{eq:heterotic_magnetic_scalar_pert}
    &\Box \Omega  = -20\tau\left[\frac{1}{2}\varphi+\Box_8 D + C - 4 A\right]\mcomma \\
    &\Box\varphi  = \frac{15}{2}\tau \varphi+5\tau  C +\frac{1}{4}\Box_2 \Omega \mcomma \\
    &\Box A  = \tau A -\frac{1}{8}\Box_2 \Omega -\frac{5}{2}\tau C-\frac{5}{4}\tau \varphi\mcomma\\
    &\Box_2 D  = 3A+C \mcomma \\
    &\Box C  = \frac{15}{4}\tau \varphi+\frac{17}{2}\tau  C +\frac{7}{8}\Box_2 \Omega \mcomma \\
    &\Box_8 D   = 4A\mcomma \\
    &0  = C+7A-\tau D-\frac{1}{2}\Omega\mperiod
\end{eqaed}

When $l=0$, there is no dependence on the internal coordinates, so that $\Omega$ and $D$ are absent. The fourth of \cref{eq:heterotic_magnetic_scalar_pert} becomes $C=-3A$. One is thus left with a $2\times 2$ matrix for $A$ and $\varphi$, whose eigenvalues are both positive.

When $l>0$, one can find two algebraic relations for $A$ and $D$ in terms of the other fields, which reduce the scalar perturbations to
\begin{eqaed}\label{eq:heterotic_scalar_mass_matrix}
    \Box_8 \begin{pmatrix}C \\ \Omega \\ \varphi \end{pmatrix} =
    \begin{pmatrix}
    L+\frac{17}{2} & -\frac{7}{8}L & \frac{15}{4}  \\
    -20 & L & -10 \\
    5 & -\frac{1}{4}L & L+\frac{15}{2}  \\
    \end{pmatrix} \tau \begin{pmatrix} C \\ \Omega \\ \varphi \end{pmatrix}\mperiod
\end{eqaed}
All the resulting eigenvalues lie above the B-F bound, but can one conclude that these heterotic vacua are stable?
There is actually a subtlety, since these vacua differ from those examined in~\cite{Basile:2018irz} due to the presence of non-abelian fields. 
Instabilities have long been known to arise, in flat space, for non-abelian gauge fields in the presence of strong electric or magnetic backgrounds~\cite{Chang:1979tg, Sikivie:1979bq}. 
These indications point indeed to the emergence of instabilities in our vacua when $H R_{S^2}^2 =\mathcal{O}(1) $. Since the product is of order $e^{-\phi_0}$, the phenomenon lies well within the corner of string theory that is well-captured by the low energy analysis, and is thus expected to play a role in our vacua. The detailed analysis of instabilities of this type in $AdS\times S$ backgrounds is an interesting problem in its own right, which is currently under investigation, and we plan to return to it in a future work.

\section{Higher derivative corrections}\label{sec:heterotic_higher_der}

These two types of classical vacua have no curvature singularities, and thus one could naively expect that they provide a complete solution of the string equations. However, the $\alpha '$ mixing induced by the tadpole deserves further consideration.

In the effective action that we studied, we included only two-derivative contributions that are of first order in $\alpha '$. Our analysis requires that both higher derivative and higher $\alpha '$ terms be negligible in the regime of validity of the vacua. We shall not focus on a specific $\alpha '$ order, or on particular higher derivative terms. Instead, simple considerations suffice to emphasize the relevant points. 

Let us begin with the heterotic model. In the string frame, the classical vacuum of \cref{eq:magnetic_vacuum_string_frame} has
\begin{eqaed}
    \frac{1}{l^2_{AdS_8}}\sim \frac{1}{R^2_{S^2}}\sim e^{2\phi_0} (\alpha')^{-1}\mcomma \qquad F^2\sim e^{2\phi_0} (\alpha ' )^{-2}\mperiod
\end{eqaed}
All $\alpha'$ corrections to \cref{eq:heterotic_action_string_frame} must be accompanied by suitable powers of $\alpha'$ for dimensional reasons. In particular, they must enter the action in \cref{eq:heterotic_action_string_frame} as
\begin{eqaed}\label{eq:higher_curvatures}
    (\alpha')^{n-1}R^{n}\mcomma
\end{eqaed}
where $R$ can be either a metric or gauge curvature. 
The contributions to the equations of motion that we considered are accompanied by $e^{2\phi_0} (\alpha ')^{-1}$. On the other hand, the higher derivative terms from \cref{eq:higher_curvatures} are accompanied by $e^{k\phi_0}(\alpha ')^{-1}$ with $k>2$. Therefore, the magnetovac solution is reliable provided that the string coupling is small enough, or, equivalently, when the magnetic flux $N_m$ is large enough.

Similar considerations apply to the orientifold vacua, for which \cref{eq:higher_curvatures} still holds. However, in this case
\begin{eqaed}
    \frac{1}{l^2_{AdS_2}}\sim \frac{1}{R^2_{S^8}}\sim e^{\phi_0} (\alpha')^{-1}\mcomma \qquad F^2\sim (\alpha ' )^{-2}\mcomma
\end{eqaed}
and, while the metric curvature terms are subleading in an $e^{\phi_0}$ series, higher derivative terms with arbitrary numbers of field strengths make the analysis not manifestly under control.

\section{Conclusions}\label{sec:conclusions}

In this paper, we have analyzed two vacua with constant dilaton profiles for the tachyon-free non-supersymmetric string models in ten dimensions. 
They are supported by constant electric or magnetic fields, and include $AdS_2$ and $AdS_8$ spacetimes. The former choice has perturbatively unstable modes in the singlet scalar sector, in close analogy~\cite{Basile:2018irz} with the vacua supported by R-R form fluxes of~\cite{Gubser:2001zr, Mourad:2016xbk}. No instability of this type is present in the $AdS_8$ case. However, other unstable modes are induced, in flat space, by non-abelian couplings in the presence of strong electric or magnetic backgrounds~\cite{Chang:1979tg, Sikivie:1979bq}. The $AdS$ counterpart of this problem is currently under investigation, but all our vacua appear to suffer from these instabilities, since the transition occurs well within regions of parameter space that are captured by the field theory analysis.

Our solutions rest on $\alpha '$ corrections to the equations of motion, and the expansion is under control only when some other parameter, the string coupling, is appropriately tuned. 
Note that $\alpha '$ corrections have been used in the past in order to break supersymmetry. For instance in~\cite{Maxfield:2014wea}, where non-trivial gauge configurations on internal tori were employed to argue for non-supersymmetric domain walls.\footnote{It would be interesting to apply the techniques of that work to our models, writing the string-frame dilaton equation as $\nabla^2 e^{-2\phi}=$ source and using the self-duality of the $F$ profile in order to satisfy a condition like \cref{eq:maxwell_propto_metric}.}
Our idea is somewhat different, since we compensate the supersymmetry breaking scalar potential, already present in the models of interest, with $\alpha'$ corrections.

Similar constant dilaton vacua should be available for T-duals of the orientifold models, for which there is no spacetime-filling dilaton potential, but the tadpole is localized on branes that break supersymmetry.
Non-trivial gauge configurations on the branes could also improve the problems related to the unbalanced tension~\cite{Gibbons:2000tf, Blumenhagen:2000dc, Dudas:2002dg}, but the complications outlined in \cref{sec:heterotic_higher_der} would remain, because the localized tadpole would still be compensated by $\alpha '$ corrections on the branes.

\section*{Acknowledgements}

We would like to thank A. Sagnotti for closely following the development of this work and for comments on the manuscript. We also thank I. Basile and S. Sethi for stimulating discussions.
This work was supported in part by Scuola Normale, by INFN (IS GSS-Pi) and by the MIUR-PRIN contract 2017CC72MK\_003.

\bibliographystyle{elsarticle-num}
\bibliography{mybib}

\begin{thebibliography}{10}
\expandafter\ifx\csname url\endcsname\relax
  \def\url#1{\texttt{#1}}\fi
\expandafter\ifx\csname urlprefix\endcsname\relax\def\urlprefix{URL }\fi
\expandafter\ifx\csname href\endcsname\relax
  \def\href#1#2{#2} \def\path#1{#1}\fi

\bibitem{AlvarezGaume:1986jb}
L.~Alvarez-Gaume, P.~H. Ginsparg, G.~W. Moore, C.~Vafa, {An O(16) x O(16)
  Heterotic String}, Phys. Lett. B171 (1986) 155--162.
\newblock \href {https://doi.org/10.1016/0370-2693(86)91524-8}
  {\path{doi:10.1016/0370-2693(86)91524-8}}.

\bibitem{Dixon:1986iz}
L.~J. Dixon, J.~A. Harvey, {String Theories in Ten-Dimensions Without
  Space-Time Supersymmetry}, Nucl. Phys. B274 (1986) 93--105, [,93(1986)].
\newblock \href {https://doi.org/10.1016/0550-3213(86)90619-X}
  {\path{doi:10.1016/0550-3213(86)90619-X}}.

\bibitem{Sagnotti:1987tw}
A.~Sagnotti, {Open Strings and their Symmetry Groups}, in: {NATO Advanced
  Summer Institute on Nonperturbative Quantum Field Theory (Cargese Summer
  Institute) Cargese, France, July 16-30, 1987}, 1987, pp. 521--528.
\newblock \href {http://arxiv.org/abs/hep-th/0208020}
  {\path{arXiv:hep-th/0208020}}.

\bibitem{Pradisi:1988xd}
G.~Pradisi, A.~Sagnotti, {Open String Orbifolds}, Phys. Lett. B216 (1989)
  59--67.
\newblock \href {https://doi.org/10.1016/0370-2693(89)91369-5}
  {\path{doi:10.1016/0370-2693(89)91369-5}}.

\bibitem{Horava:1989vt}
P.~Horava, {Strings on World Sheet Orbifolds}, Nucl. Phys. B327 (1989)
  461--484.
\newblock \href {https://doi.org/10.1016/0550-3213(89)90279-4}
  {\path{doi:10.1016/0550-3213(89)90279-4}}.

\bibitem{Horava:1989ga}
P.~Horava, {Background Duality of Open String Models}, Phys. Lett. B231 (1989)
  251--257.
\newblock \href {https://doi.org/10.1016/0370-2693(89)90209-8}
  {\path{doi:10.1016/0370-2693(89)90209-8}}.

\bibitem{Bianchi:1990yu}
M.~Bianchi, A.~Sagnotti, {On the systematics of open string theories}, Phys.
  Lett. B247 (1990) 517--524.
\newblock \href {https://doi.org/10.1016/0370-2693(90)91894-H}
  {\path{doi:10.1016/0370-2693(90)91894-H}}.

\bibitem{Bianchi:1990tb}
M.~Bianchi, A.~Sagnotti, {Twist symmetry and open string Wilson lines}, Nucl.
  Phys. B361 (1991) 519--538.
\newblock \href {https://doi.org/10.1016/0550-3213(91)90271-X}
  {\path{doi:10.1016/0550-3213(91)90271-X}}.

\bibitem{Bianchi:1991eu}
M.~Bianchi, G.~Pradisi, A.~Sagnotti, {Toroidal compactification and symmetry
  breaking in open string theories}, Nucl. Phys. B376 (1992) 365--386.
\newblock \href {https://doi.org/10.1016/0550-3213(92)90129-Y}
  {\path{doi:10.1016/0550-3213(92)90129-Y}}.

\bibitem{Sagnotti:1992qw}
A.~Sagnotti, {A Note on the Green-Schwarz mechanism in open string theories},
  Phys. Lett. B294 (1992) 196--203.
\newblock \href {http://arxiv.org/abs/hep-th/9210127}
  {\path{arXiv:hep-th/9210127}}, \href
  {https://doi.org/10.1016/0370-2693(92)90682-T}
  {\path{doi:10.1016/0370-2693(92)90682-T}}.

\bibitem{Sagnotti:1995ga}
A.~Sagnotti, {Some properties of open string theories}, in: {Supersymmetry and
  unification of fundamental interactions. Proceedings, International Workshop,
  SUSY 95, Palaiseau, France, May 15-19}, 1995, pp. 473--484.
\newblock \href {http://arxiv.org/abs/hep-th/9509080}
  {\path{arXiv:hep-th/9509080}}.

\bibitem{Sagnotti:1996qj}
A.~Sagnotti, {Surprises in open string perturbation theory}, Nucl. Phys. Proc.
  Suppl. 56B (1997) 332--343.
\newblock \href {http://arxiv.org/abs/hep-th/9702093}
  {\path{arXiv:hep-th/9702093}}, \href
  {https://doi.org/10.1016/S0920-5632(97)00344-7}
  {\path{doi:10.1016/S0920-5632(97)00344-7}}.

\bibitem{Sugimoto:1999tx}
S.~Sugimoto, {Anomaly cancellations in type I D-9 - anti-D-9 system and the
  USp(32) string theory}, Prog. Theor. Phys. 102 (1999) 685--699.
\newblock \href {http://arxiv.org/abs/hep-th/9905159}
  {\path{arXiv:hep-th/9905159}}, \href {https://doi.org/10.1143/PTP.102.685}
  {\path{doi:10.1143/PTP.102.685}}.

\bibitem{Antoniadis:1999xk}
I.~Antoniadis, E.~Dudas, A.~Sagnotti, {Brane supersymmetry breaking}, Phys.
  Lett. B 464 (1999) 38--45.
\newblock \href {http://arxiv.org/abs/hep-th/9908023}
  {\path{arXiv:hep-th/9908023}}, \href
  {https://doi.org/10.1016/S0370-2693(99)01023-0}
  {\path{doi:10.1016/S0370-2693(99)01023-0}}.

\bibitem{Angelantonj:1999jh}
C.~Angelantonj, {Comments on open string orbifolds with a nonvanishing B(ab)},
  Nucl. Phys. B 566 (2000) 126--150.
\newblock \href {http://arxiv.org/abs/hep-th/9908064}
  {\path{arXiv:hep-th/9908064}}, \href
  {https://doi.org/10.1016/S0550-3213(99)00662-8}
  {\path{doi:10.1016/S0550-3213(99)00662-8}}.

\bibitem{Aldazabal:1999jr}
G.~Aldazabal, A.~M. Uranga, {Tachyon free nonsupersymmetric type IIB
  orientifolds via Brane - anti-brane systems}, JHEP 10 (1999) 024.
\newblock \href {http://arxiv.org/abs/hep-th/9908072}
  {\path{arXiv:hep-th/9908072}}, \href
  {https://doi.org/10.1088/1126-6708/1999/10/024}
  {\path{doi:10.1088/1126-6708/1999/10/024}}.

\bibitem{Angelantonj:1999ms}
C.~Angelantonj, I.~Antoniadis, G.~D'Appollonio, E.~Dudas, A.~Sagnotti, {Type I
  vacua with brane supersymmetry breaking}, Nucl. Phys. B 572 (2000) 36--70.
\newblock \href {http://arxiv.org/abs/hep-th/9911081}
  {\path{arXiv:hep-th/9911081}}, \href
  {https://doi.org/10.1016/S0550-3213(00)00052-3}
  {\path{doi:10.1016/S0550-3213(00)00052-3}}.

\bibitem{Angelantonj:2002ct}
C.~Angelantonj, A.~Sagnotti, {Open strings}, Phys. Rept. 371 (2002) 1--150,
  [Erratum: Phys.Rept. 376, 407 (2003)].
\newblock \href {http://arxiv.org/abs/hep-th/0204089}
  {\path{arXiv:hep-th/0204089}}, \href
  {https://doi.org/10.1016/S0370-1573(02)00273-9}
  {\path{doi:10.1016/S0370-1573(02)00273-9}}.

\bibitem{Mourad:2017rrl}
J.~Mourad, A.~Sagnotti, {An Update on Brane Supersymmetry Breaking} (11 2017).
\newblock \href {http://arxiv.org/abs/1711.11494} {\path{arXiv:1711.11494}}.

\bibitem{Basile:2020xwi}
I.~Basile, {On String Vacua without Supersymmetry: brane dynamics, bubbles and
  holography}, Ph.D. thesis, Pisa, Scuola Normale Superiore (2020).
\newblock \href {http://arxiv.org/abs/2010.00628} {\path{arXiv:2010.00628}}.

\bibitem{Basile:2021mkd}
I.~Basile, {Supersymmetry breaking, brane dynamics and Swampland conjectures},
  JHEP 10 (2021) 080.
\newblock \href {http://arxiv.org/abs/2106.04574} {\path{arXiv:2106.04574}},
  \href {https://doi.org/10.1007/JHEP10(2021)080}
  {\path{doi:10.1007/JHEP10(2021)080}}.

\bibitem{Basile:2021vxh}
I.~Basile, {Supersymmetry breaking and stability in string vacua: Brane
  dynamics, bubbles and the swampland}, Riv. Nuovo Cim. 44~(10) (2021)
  499--596.
\newblock \href {http://arxiv.org/abs/2107.02814} {\path{arXiv:2107.02814}},
  \href {https://doi.org/10.1007/s40766-021-00024-9}
  {\path{doi:10.1007/s40766-021-00024-9}}.

\bibitem{Dudas:2000ff}
E.~Dudas, J.~Mourad, {Brane solutions in strings with broken supersymmetry and
  dilaton tadpoles}, Phys. Lett. B 486 (2000) 172--178.
\newblock \href {http://arxiv.org/abs/hep-th/0004165}
  {\path{arXiv:hep-th/0004165}}, \href
  {https://doi.org/10.1016/S0370-2693(00)00734-6}
  {\path{doi:10.1016/S0370-2693(00)00734-6}}.

\bibitem{Gubser:2001zr}
S.~S. Gubser, I.~Mitra, {Some interesting violations of the
  Breitenlohner-Freedman bound}, JHEP 07 (2002) 044.
\newblock \href {http://arxiv.org/abs/hep-th/0108239}
  {\path{arXiv:hep-th/0108239}}, \href
  {https://doi.org/10.1088/1126-6708/2002/07/044}
  {\path{doi:10.1088/1126-6708/2002/07/044}}.

\bibitem{Mourad:2016xbk}
J.~Mourad, A.~Sagnotti, {$AdS$ Vacua from Dilaton Tadpoles and Form Fluxes},
  Phys. Lett. B 768 (2017) 92--96.
\newblock \href {http://arxiv.org/abs/1612.08566} {\path{arXiv:1612.08566}},
  \href {https://doi.org/10.1016/j.physletb.2017.02.053}
  {\path{doi:10.1016/j.physletb.2017.02.053}}.

\bibitem{Freedman:1983xa}
D.~Z. Freedman, G.~W. Gibbons, {Electrovac Ground State in Gauged SU(2) X SU(2)
  Supergravity}, Nucl. Phys. B 233 (1984) 24--49.
\newblock \href {https://doi.org/10.1016/0550-3213(84)90168-8}
  {\path{doi:10.1016/0550-3213(84)90168-8}}.

\bibitem{Antoniadis:1989mn}
I.~Antoniadis, C.~Bachas, A.~Sagnotti, {Gauged Supergravity Vacua in String
  Theory}, Phys. Lett. B 235 (1990) 255--260.
\newblock \href {https://doi.org/10.1016/0370-2693(90)91960-J}
  {\path{doi:10.1016/0370-2693(90)91960-J}}.

\bibitem{Breitenlohner:1982jf}
P.~Breitenlohner, D.~Z. Freedman, {Stability in Gauged Extended Supergravity},
  Annals Phys. 144 (1982) 249.
\newblock \href {https://doi.org/10.1016/0003-4916(82)90116-6}
  {\path{doi:10.1016/0003-4916(82)90116-6}}.

\bibitem{Basile:2018irz}
I.~Basile, J.~Mourad, A.~Sagnotti, {On Classical Stability with Broken
  Supersymmetry}, JHEP 01 (2019) 174.
\newblock \href {http://arxiv.org/abs/1811.11448} {\path{arXiv:1811.11448}},
  \href {https://doi.org/10.1007/JHEP01(2019)174}
  {\path{doi:10.1007/JHEP01(2019)174}}.

\bibitem{Chang:1979tg}
S.-J. Chang, N.~Weiss, {Instability of Constant {Yang-Mills} Fields}, Phys.
  Rev. D 20 (1979) 869.
\newblock \href {https://doi.org/10.1103/PhysRevD.20.869}
  {\path{doi:10.1103/PhysRevD.20.869}}.

\bibitem{Sikivie:1979bq}
P.~Sikivie, {Instability of Abelian Field Configurations in {Yang-Mills}
  Theory}, Phys. Rev. D 20 (1979) 877.
\newblock \href {https://doi.org/10.1103/PhysRevD.20.877}
  {\path{doi:10.1103/PhysRevD.20.877}}.

\bibitem{Maxfield:2014wea}
T.~Maxfield, S.~Sethi, {Domain Walls, Triples and Acceleration}, JHEP 08 (2014)
  066.
\newblock \href {http://arxiv.org/abs/1404.2564} {\path{arXiv:1404.2564}},
  \href {https://doi.org/10.1007/JHEP08(2014)066}
  {\path{doi:10.1007/JHEP08(2014)066}}.

\bibitem{Gibbons:2000tf}
G.~W. Gibbons, R.~Kallosh, A.~D. Linde, {Brane world sum rules}, JHEP 01 (2001)
  022.
\newblock \href {http://arxiv.org/abs/hep-th/0011225}
  {\path{arXiv:hep-th/0011225}}, \href
  {https://doi.org/10.1088/1126-6708/2001/01/022}
  {\path{doi:10.1088/1126-6708/2001/01/022}}.

\bibitem{Blumenhagen:2000dc}
R.~Blumenhagen, A.~Font, {Dilaton tadpoles, warped geometries and large extra
  dimensions for nonsupersymmetric strings}, Nucl. Phys. B 599 (2001) 241--254.
\newblock \href {http://arxiv.org/abs/hep-th/0011269}
  {\path{arXiv:hep-th/0011269}}, \href
  {https://doi.org/10.1016/S0550-3213(01)00028-1}
  {\path{doi:10.1016/S0550-3213(01)00028-1}}.

\bibitem{Dudas:2002dg}
E.~Dudas, J.~Mourad, C.~Timirgaziu, {Time and space dependent backgrounds from
  nonsupersymmetric strings}, Nucl. Phys. B 660 (2003) 3--24.
\newblock \href {http://arxiv.org/abs/hep-th/0209176}
  {\path{arXiv:hep-th/0209176}}, \href
  {https://doi.org/10.1016/S0550-3213(03)00248-7}
  {\path{doi:10.1016/S0550-3213(03)00248-7}}.

\end{thebibliography}

\end{document}